\documentstyle[11pt,epsfig,a4]{article}
\begin{document}

\def\Journal#1#2#3#4{{#1} {\bf #2}, #3 (#4)}

\def\etal{{\it et\ al.}}
\def\NCA{\em Nuovo Cim.}
\def\NIM{\em Nucl. Instrum. Methods}
\def\NIMA{{\em Nucl. Instrum. Methods} A}
\def\NPB{{\em Nucl. Phys.} B}
\def\PLB{{\em Phys. Lett.}  B}
\def\PRL{\em Phys. Rev. Lett.}
\def\PRC{{\em Phys. Rev.} C}
\def\PRD{{\em Phys. Rev.} D}
\def\ZPC{{\em Z. Phys.} C}
\def\ASP{{\em Astrop. Phys.}}
\def\JETP{{\em JETP Lett.\ }}

\def\numunue{\nu_\mu\rightarrow\nu_e}
\def\numunutau{\nu_\mu\rightarrow\nu_\tau}
\def\nuebar{\bar\nu_e}
\def\numubar{\bar\nu_\mu}
\def\numubarnuebar{\bar\nu_\mu\rightarrow\bar\nu_e}

\title{Long baseline neutrino oscillation disappearance search \\
using a $\nu$ beam from muon decays}
\author{A. Bueno, M. Campanelli, A. Rubbia\footnote{On leave from CERN, Geneva, Switzerland.}\\
{\it Institut f\"{u}r Teilchenphysik} \\ 
{\it ETHZ, CH-8093 Z\"{u}rich, Switzerland}}
\maketitle
\abstract{We study the feasibility of performing a $\nu_\mu$ disappearance 
long-baseline experiment using a neutrino beam coming from muon decays.
The advantage of such a technique with respect to
the production of neutrino beams from pions is that in a muon decay both muon
and electron neutrinos are produced in the same quantity.
In case of $\nu_\mu\to\nu_{\tau,s}$ oscillations, the $\nu_e$ charged current
(CC) events can 
be used as a control sample, to predict in situ the $\nu_\mu$ rates, 
thus reducing the systematics due to the knowledge of the 
neutrino flux, which is the main source of uncertainties for disappearance
experiments. We consider as our neutrino target, the ICARUS detector 
in its final mass configuration of 4.8 kton.}
\vspace{1cm}
\section{Introduction}
The most satisfactory explanation of the observed disappearance of 
atmospheric muon neutrinos result, strongly confirmed 
by the latest observations of the Super-Kamiokande\cite{sk} and SoudanII 
\cite{soudan} collaborations, is
$\nu_\mu-\nu_\tau$ (or $\nu_\mu-\nu_s$) oscillations.\par
The importance of this result requires however confirmation in 
a different environment, i.e. using neutrinos produced from accelerators.\par
Several proposals have been made to test the oscillations using a long 
baseline experiment, in Europe \cite{ica,icacernmi,opera,noe,aquarich}, 
in Japan \cite{jap}, and in the 
US \cite{us}.
The first long-baseline experiment, K2K, will start data taking in January 
1999.
This experiment, based on a flux of muon neutrinos from the KEK laboratory to
the Kamioka mine, is expected to test $\nu_\mu\to\nu_X$ disappearance reaching
the region down to 
$\Delta m^2=3\times10^{-3} eV^2$ at 90\% C.L. after three years of running.\par
Two of the proposed experiments for the CERN-LNGS long-baseline project 
(ICARUS\cite{ica} and OPERA \cite{opera}) will be able to observe the
appearance of the $\tau$ in $\numunutau$ oscillation, and will ultimately 
reach the sensitivity of $\Delta m^2$ about $1.5\times 10^{-3} eV^2$ at 90\%
C.L. \cite{pietropaolo}.\par
While the region for oscillations suggested by the SuperKamiokande result
is quite precise in the determination of the mixing angle consistent with
maximal mixing, the
uncertainty in $\Delta m^2$ indicated by the SuperKamiokande zenith angle 
distribution is quite large, and does not exclude the possibility
of this parameter to be as small as $6\times10^{-4} eV^2$. For this reason it 
is  important to envisage an accelerator-based experiment able to cover the
full parameter space.\par
The probability for neutrino oscillations between two families is given by 
\begin{eqnarray}
P(E_\nu)=\sin^2 2\theta \sin^2(1.27 \Delta m^2 \frac{L}{E_\nu})
\label{eq:pro1}
\end{eqnarray}
where $\theta$ is the mixing angle between the two neutrino flavours, and
$\Delta m^2$ (in $eV^2$) is the difference of the squares of the neutrino masses.
In the region with full mixing and small probability, it can be approximated as
\begin{eqnarray}
P(E_\nu)\approx(1.27 \Delta m^2 L/E_\nu)^2
\label{eq:approb}
\end{eqnarray}
Thus, using a baseline of 732 km and neutrino energies of a few GeV, in the 
region suggested by the atmospheric neutrinos the oscillation probability is
of the order of a few per cent.\par
The ultimate sensitivity at low $\Delta m^2$ in the $\nu_\mu\to\nu_X$ 
disappearance mode will
be limited by the knowledge of the neutrino flux. In fact, since the 
oscillation probability shows a $1/E^2$ dependence, the oscillated events are
all concentrated in the low-energy region, where it is very difficult to 
observe any shape distorsion due to the oscillation. So, the only visible 
effect is a
deficit in the total number of neutrinos observed. The integrated oscillation
probability in this case is
\begin{eqnarray}
{\cal P}=\frac{\int P(E) \Phi_{\nu_\mu}(E)dE}{\int \Phi_{\nu_\mu}(E)dE}\approx
(\Delta m^2 L)^2\frac{\int\Phi_{\nu_\mu}(E)/E^2 dE}{\int \Phi_{\nu_\mu}(E)dE}
\label{eq:pro}
\end{eqnarray}
showing a quadratic dependence on $\Delta m^2$.\par
In long-baseline experiments performed using a pion beam, the problem of the 
flux uncertainty has been
addressed by the use of a near detector, this solution having the disadvantage
that the solid angle covered by the two detectors is not the same, thus 
requiring a 
non-trivial extrapolation from the near to the far flux\cite{minos}
\cite{nice}.\par
In the last few years there has been growing interest 
in studying the technical feasibility and the impact on physics of a 
muon collider with high intensity beams\cite{muonco}. 
Muons decaying in this machine are a
continuous source of both muon and electron-like neutrinos \cite{geer}
via the following relation\footnote{In this paper we always refer to the 
decay of negative muons. The same considerations are obviously valid in the
case of positive muons.}:
\begin{eqnarray}
\mu^- \rightarrow e^-\bar\nu_e\nu_\mu.
\end{eqnarray}
Between production and detection, neutrinos oscillate with a certain
probability. Let us 
consider three possible cases of mixing between two neutrino flavours:
\begin{enumerate}
\item Pure {\bf$\numunutau$} oscillations: in this case there will be 
disappearance of $\nu_\mu$ neutrinos, with oscillation probability P, given
by equation \ref{eq:pro1}. If we define as $\Phi_{\nu_\mu}^i$ the initial
$\nu_\mu$ flux, the oscillated flux $\Phi_{\nu_\mu}$ will be:
\begin{eqnarray}
\Phi_{\nu_\mu}=\Phi_{\nu_\mu}^i\times (1-P)
\end{eqnarray}
The $\nu_\tau$ produced by the oscillation will have a flux $\Phi_{\nu_\tau}$
\begin{eqnarray}
\Phi_{\nu_\tau}=\Phi_{\nu_\mu}^i\times P
\end{eqnarray}
but will not be seen in the detector because of the kinematical suppression 
resulting from the $\tau$ mass.
The $\bar\nu_e$ component of the beam remains the same, since in this case
neutrino electrons do not oscillate. The $\bar{\nu_e}$ beam, however,
provides a way to predict at the far detector the initial $\nu_\mu$ flux. 
In the muon rest frame, the distribution of muon neutrinos and electron 
antineutrinos is precisely predicted by the V-A theory:
\begin{eqnarray}
\frac{d^2N_{\nu_\mu}}{dxd\Omega}\propto\frac{12x^2}{4\pi}[(3-2x)+(1-2x)P_\mu
\cos\theta]\\
\frac{d^2N_{\bar{\nu_e}}}{dxd\Omega}\propto\frac{12x^2}{4\pi}[(1-x)+(1-x)P_\mu
\cos\theta]
\end{eqnarray}
where $x\equiv 2E_\nu/m_\mu$, $P_\mu$ is the average polarization of the muon 
beam, and $\theta$ is the angle between the momentum vector of the neutrino 
and the mean angle of the muon polarization. In a muon collider, the 
beam polarization can be carefully measured via the spectrum of the electrons
from muon decay; thus the spectra of the two components of the neutrino
beam can be known with good accuracy, and so the ratio of
the fluxes reaching the far detector .
\item Pure {\bf$\nu_\mu\to\nu_s$} oscillations: the $\nu_\mu$ flux and 
spectrum
are exactly the same as for the $\numunutau$ case, and also the $\bar{\nu_e}$
beam is not affected. It is possible, however, to discriminate between
$\nu_\mu\to\nu_s$ and $\numunutau$ oscillations, since the ratio of neutral 
current to charged interactions is different in the two cases.\par
\item Pure {\bf$\numunue$} oscillations: in this case, also electron 
antineutrinos would oscillate into $\bar{\nu_\mu}$, with the same probability
as the $\numunue$ oscillation. If we call this probability P (given as above
by equation \ref{eq:pro1}), the final flux will be composed of four neutrino
types:
\begin{eqnarray}
\Phi_{\nu_\mu}& = &\Phi_{\nu_\mu}^i\times (1-P)\\
\Phi_{\bar{\nu_\mu}}& = &\Phi_{\bar{\nu_e}}^i\times P\\
\Phi_{\nu_e}& = &\Phi_{\nu_\mu}^i\times P\\
\Phi_{\bar{\nu_e}}& = &\Phi_{\bar{\nu_e}}^i\times (1-P)
\end{eqnarray}
The total detected flux $\Phi_{\nu_\mu}+\Phi_{\bar{\nu_\mu}}+\Phi_{\nu_e}+
\Phi_{\bar{\nu_e}}$ in this case would be the same, i.e. equal to the
initial flux.
The antineutrinos coming from the oscillation would produce leptons of opposite
sign with respect to neutrinos from the beam, so
a magnetic detector with charge discrimination capability would be able to
study in detail this kind of processes, as explained in \cite{sbm} for a 
different choice of the oscillation parameters.
If no magnetic detector is available, it is possible to use the
difference between neutrinos and antineutrinos cross-sections to get some hint
for this kind of process. Let us call R the ratio between neutrino and 
antineutrino cross sections:
\begin{eqnarray}
R=\frac{\sigma(\nu,CC)}{\sigma(\bar{\nu},CC)}
\end{eqnarray}
The fluxes of $\bar{\nu_e}$ and $\nu_\mu$ are equal at 
production, but since the energy spectra are different, the number of neutrinos
reaching the far detector is slightly different for the two flavours. We 
assume this effect to be negligible, and we set $\Phi_{\nu_\mu}^i\approx
\Phi_{\bar{\nu_e}}^i\equiv\Phi$
(deviations from this behaviour can anyway be computed for a more accurate
estimation). Thus, the number of interactions for the different kind
neutrino flavours would be (assuming no charge discrimination):
\begin{eqnarray}
N_\mu(E)\equiv N_{\nu_\mu}+N_{\bar{\nu_\mu}}=A\times(1-P) + A/R\times P\\
N_e(E)\equiv N_{\nu_e}+N_{\bar{\nu_e}}=A\times P + A\times (1-P)/R
\end{eqnarray}
where A is the product of the flux times the neutrino cross section. The
ratio between the two flavours is still dependent on P:
\begin{eqnarray}
\frac{N_\mu}{N_e}=\frac{1-P+P/R}{P+(1-P)/R}\approx\frac{2-P}{1+P}
\end{eqnarray}
where we have assumed $R=2$, independent of energy. The above 
equation allows to derive the oscillation probability, with no assumption
in the initial neutrino flux. Without charge assignment
the interpretation of an observed deficit in terms of 
$\nu_\mu$ disappearance or $\numunue$ oscillations is not straightforward, 
and the ambiguity could be solved considering data from other experiments
(as usually done in the interpretation of the SuperKamiokande results) or
studying the modification in the $y_{bj}$ distribution due to the 
the antineutrinos produced in $\numunue$ oscillations.
\end{enumerate}
\section{Event rates}
The muons are produced in the 
decay chain of hadrons behind an appropriate target and are subsequently captured, 
cooled, accelerated and stored into a ring where they are let to decay. 

The muon yield is determined by several parameters.
If a pulsed proton source is assumed, the total number
of muons is given by the total proton yield times the average
pion yield per proton, times the muon yield per pion, so that the total number
of muons accelerated is given by the following relation:
\[N_{\mu}=f_{bunch}\times N_{p/bunch}\times Y_{\pi/p}\times Y_{\mu/\pi}\times
t\]
We assume a bunch repetition frequency of 10 Hz, with $10^{14}$ protons per 
bunch. Given a standard yearly running of $10^{7} s$, it corresponds to
$10^{22}$ protons per year.\par 
Based on results from muon collider studies\cite{muonco}, we assume that the
yield is about 0.6 pions per proton, half of which can produce useful muons,
and 0.25 muons per pion survived cooling and acceleration, thus resulting in 
about 0.1 muons per proton. Overall, a
total muon production above $10^{21}$ muons per year can be achieved.
Let us remark that the requirement on cooling should be less stringent than that
for a high luminosity muon collider where overlap between two high density beams
is required.
We assume that the total muon flux can be improved by a factor 2 with respect
to the figures quoted above. By running the experiment during 5 years, a total of
$10^{22}$ muons could be reached.
\par
\begin{figure}
  \begin{center}\mbox{
   \epsfig{file=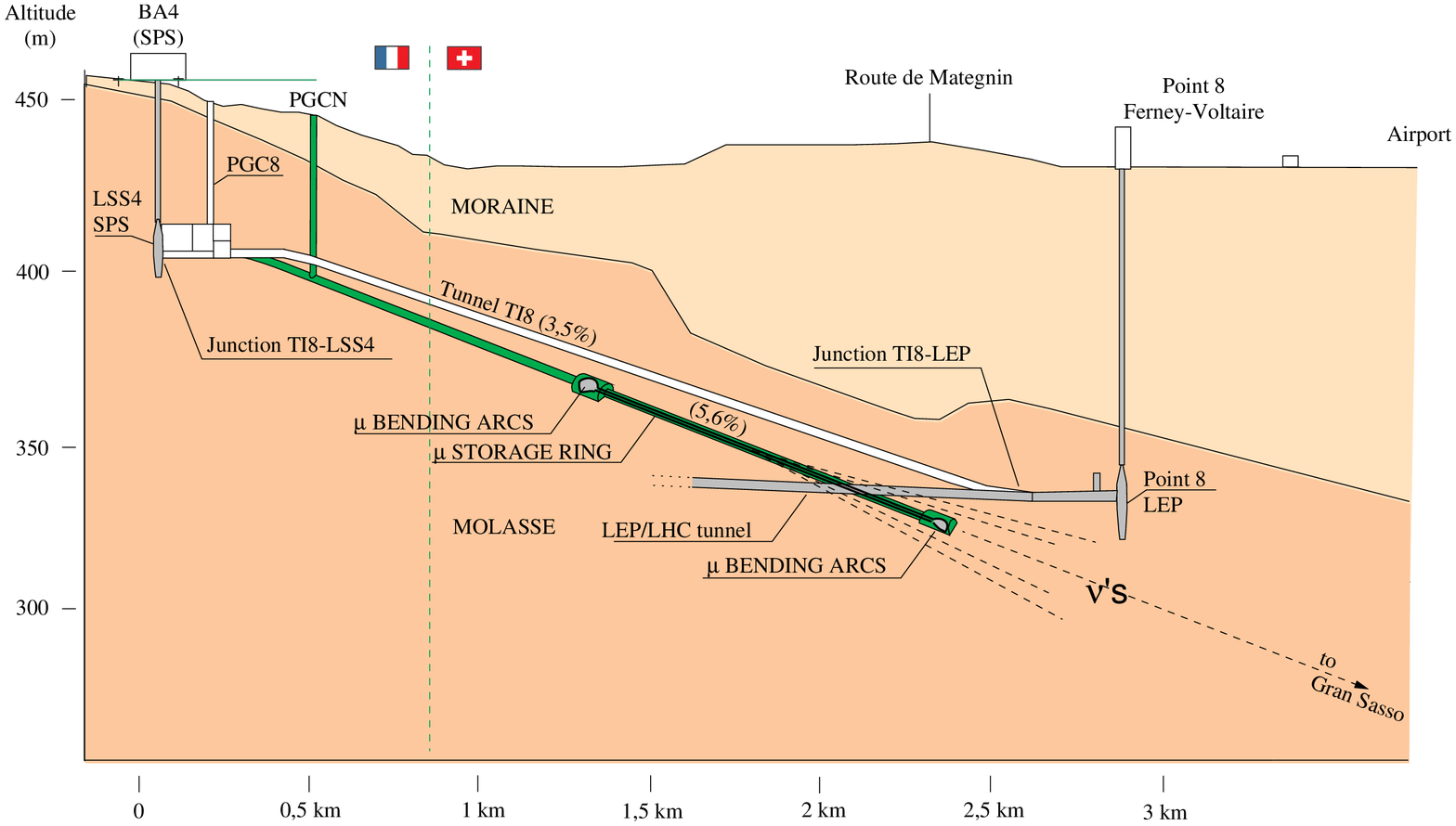,width=.95\linewidth}
}
  \end{center}
  \caption{Layout of the tunnel for the CERN-Gran Sasso beam}
  \label{fig:ngs}
\end{figure}

We assume that the storage ring would be placed inside the
proposed tunnel for the CERN-Gran Sasso neutrino facility (NGS)\cite{ngs}
(see Figure \ref{fig:ngs}).
An optimized geometrical configuration of the storage ring
is that of a ''cigare'', in which two long straight sections are closed
by two small arcs in which the muons are strongly bent. The radius of curvature
(in meters) of a charged particle in a uniform magnetic field is given by
\[R=\frac{p}{0.3 B}\]
where $p$ is the momentum normal to the magnetic field in GeV, and B is the
value of the field in Tesla. 
The straight section would fit in the decay tunnel of the NGS, which is 
foreseen to have a diameter of about 3 meters and a length of about 1 km. 
Given this configuration, the losses in the arcs can be neglected, and about 
50\%
of the muon decays are directed towards the target.\par
Using conventional magnets for the dipoles of the arcs, and assuming a magnetic
field of 2 Tesla, the radius of curvature would be about 10 meters, therefore some
modification would be needed to the proposed target and hadron stop caverns
to accomodate the arcs. If on the other hand superconducting magnets could
be used (8 Tesla), the radius
of curvature would be around 2.5 meters, and both arcs could fit in the 
caverns proposed in the NGS project.
\par
In this configuration, half of the neutrinos would be emitted in the backward
direction with respect to the Gran Sasso laboratory. Given the slope of the
tunnel, they would reach the surface in a site located about 1.5 km from 
the muon accelerator, between the CERN-Meyrin and CERN-Prevessin laboratories.
There, a detector could profit
from the extremely intense beam to provide
a further cross-check on the neutrino flux. This location would act
as a near detector, but in a much more favorable location.\par
In the other direction, the neutrinos directed towards the Gran Sasso would 
travel the distance between
the two laboratories (732 km), to be detected in one of the large apparatus  
installed there.\par
We consider the ICARUS \cite{ica} detector as a target, in its final mass
configuration of 4.8 ktons. We have there an excellent energy
resolution for electrons, and a very good electron-muon separation capability.
We purposely ignore the charge measurement, not considering
the possibility of studying $\numunue$ oscillations using the charge of the
lepton as a discriminator.
\par
\begin{figure}
  \begin{center}\mbox{
   \epsfig{file=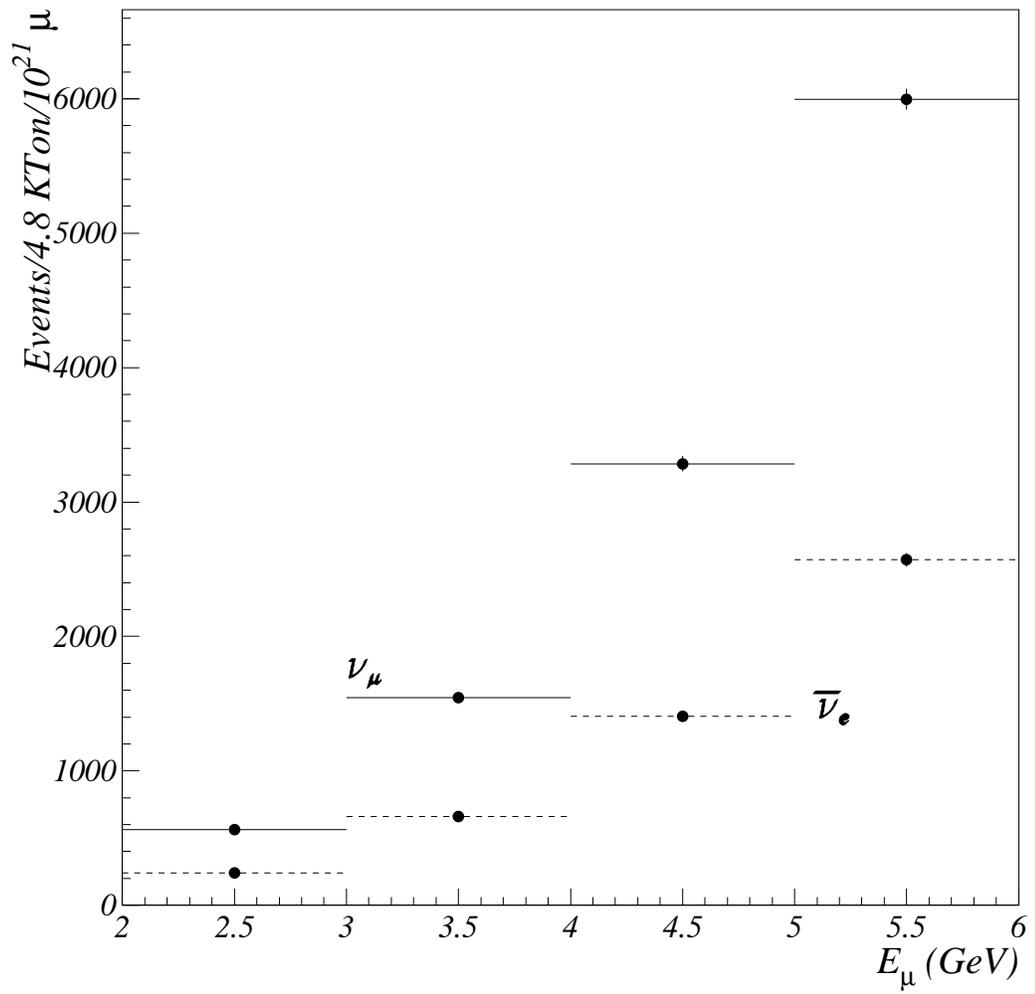,bb=0 120 567 687,width=.95\linewidth}
}
  \end{center}
  \caption{Predicted rates of $\nu_\mu$ (full line) and $\bar{\nu_e}$ (dashed line) events for $10^{21}$ muons and 4.8 kton detector, as a function of muon beam energy}
  \label{fig:rates}
\end{figure}
\begin{figure}
  \begin{center}\mbox{
   \epsfig{file=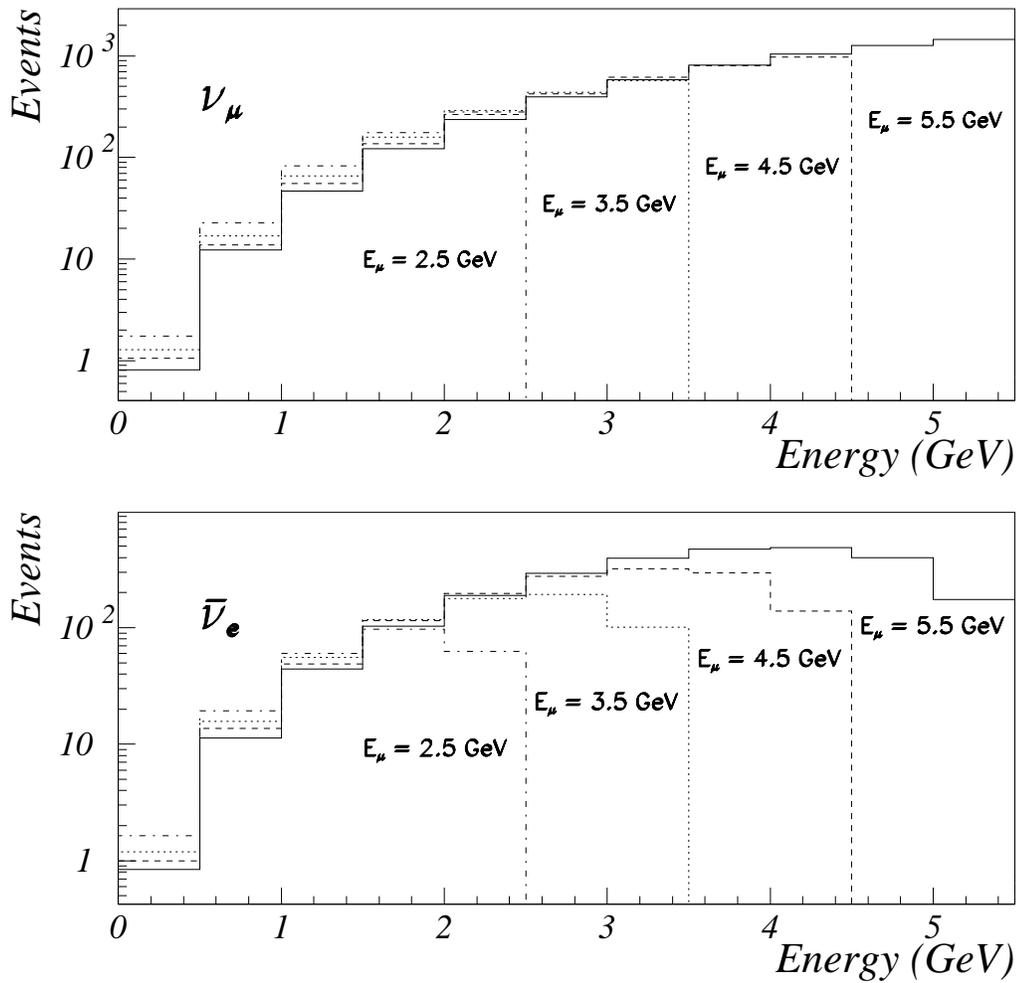,bb=0 120 567 687,width=.95\linewidth}
}
  \end{center}
  \caption{Energy spectra of $\nu_\mu$ (upper plot) and $\bar{\nu_e}$ (lower
plot) interacting via charged current processes for muon energies of 2.5, 3.5, 4.5 and 5.5 GeV ($10^{21}$ muons, 4.8 kton).}
  \label{fig:spectra}
\end{figure}

The total number of $\nu_\mu$ and $\bar{\nu_e}$ charged current interactions
in ICARUS as a function of the muon beam energy is shown in 
Figure \ref{fig:rates} and in Table \ref{tab:ratest}, considering unpolarized
muon beams.
The energy spectra of the two types of neutrinos for different values of 
the muon energy are shown in Figure \ref{fig:spectra}.
We have assumed for simplicity that the neutrino
cross sections scale linearly with energy even in the very
low-energy part of the spectrum:
\[\sigma^{CC}(\nu_e)=\sigma^{CC}(\nu_\mu)=0.67\times 10^{-38} \times E (GeV)\ \ cm^2\]
\[\sigma^{CC}(\bar{\nu_e})=\sigma^{CC}(\bar{\nu_\mu})=0.34 \times 10^{-38} \times E(GeV)\ \ cm^2\]
The relevant point is that for the energies considered here the cross
section for electron and muon neutrinos are the same.
\begin{table}
\begin{center}
\begin{tabular}{||c|r|r|}\hline \hline
Muon energy (GeV)&$\nu_\mu$ CC events&$\bar{\nu_e}$ CC events\\ \hline
2.5&560&240\\
3.5&1540&660\\
4.5&3280&1410\\
5.5&6000&2570\\ \hline
\end{tabular}
\end{center}
\caption{Charge current rates in the ICARUS detector for neutrinos produced
by the decay of $10^{21}$ muons, at different muon beam energies.}
\label{tab:ratest}
\end{table}

\section{Results}
We consider the results that could be achieved by the experimental setup 
described above, without near detector, in the case of pure 
$\nu_\mu\to\nu_{\tau,s}$ oscillation.\par
\begin{figure}
  \begin{center}\mbox{
   \epsfig{file=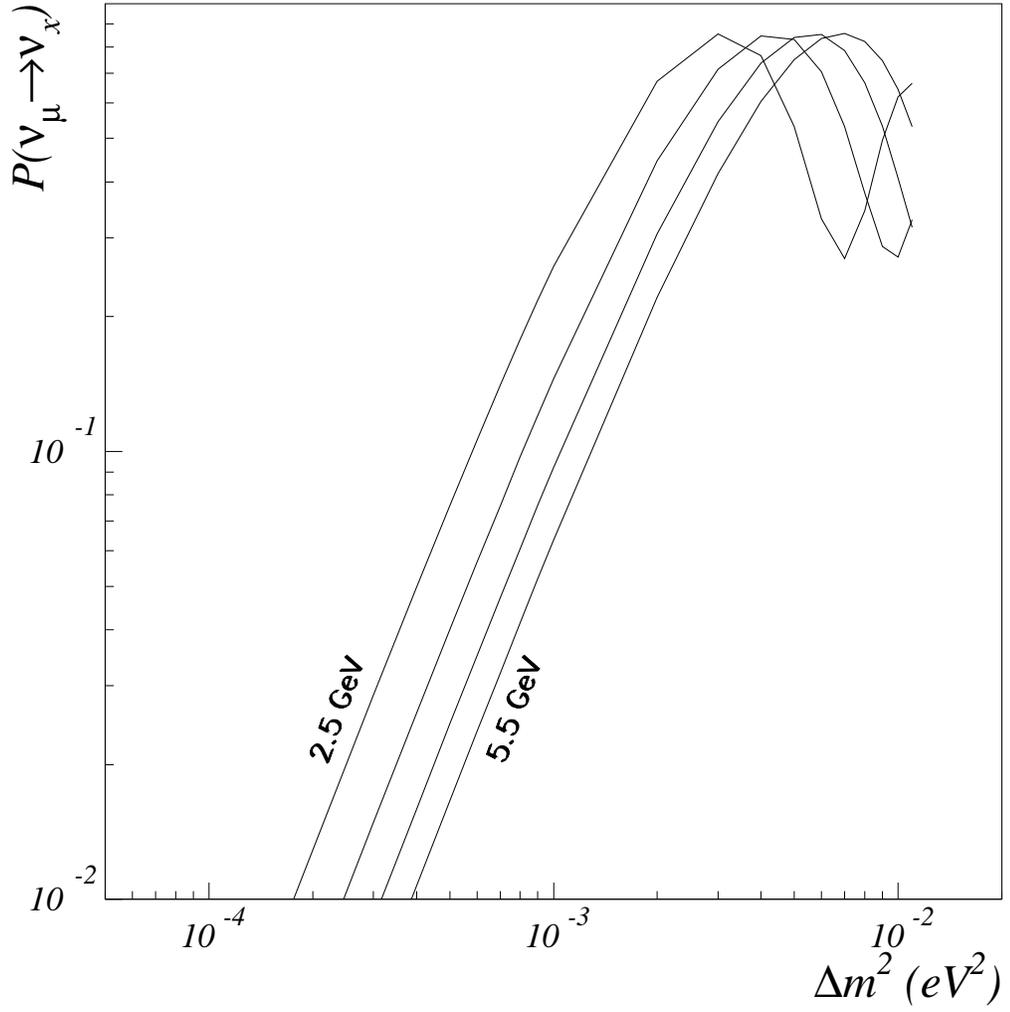,bb=0 120 567 687,width=.95\linewidth}
}
  \end{center}
  \caption{Average integrated (over the full energy spectrum) oscillation probability of $\nu_\mu$ as a function of $\Delta m^2$
for muon energies of 2.5, 3.5, 4.5 and 5.5 GeV (curves from left to right).}
  \label{fig:probosc}
\end{figure}

For the different values of muon beam energy, the oscillation probability
as a function of the parameter $\Delta m^2$ is shown in Figure 
\ref{fig:probosc}. In the region of small probabilities, the dependence of
this quantity on $\Delta m^2$ is quadratic, as seen from equation 
\ref{eq:approb}. The probability has also
a quadratic dependence on the neutrino energy,
so the low-energy muons yield a higher oscillation probability with respect to the 
high-energy muons.
In the $\Delta
m^2$ region of interest, between $10^{-4}$ and $10^{-3}\ eV^2$, the oscillation 
probability is
between 1 and 10\%, and a similar or better sensitivity is needed
for observing the oscillation signal.\par
\begin{figure}
  \begin{center}\mbox{
   \epsfig{file=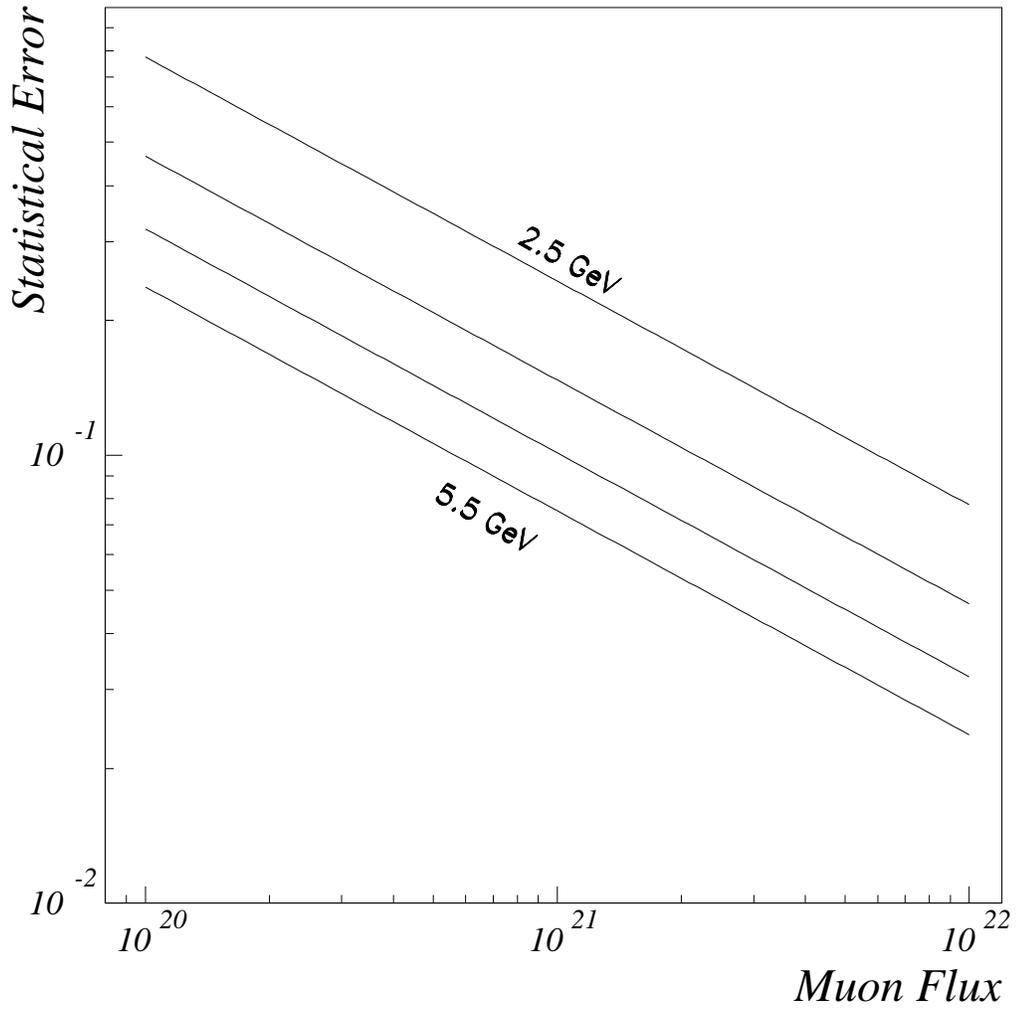,bb=0 120 567 687,width=.95\linewidth}
}
  \end{center}
  \caption{Total relative error (oscillation plus control sample) as a function of the total muon flux}
  \label{fig:errel}
\end{figure}

If the energy shape of the oscillated events is not considered, and only a 
counting experiment is performed, as is the case for the small-$\Delta m^2$
region, the statistical error is just given by the total number of muon 
neutrinos observed:
\[\frac{\Delta N_\mu}{N_\mu}=\frac{1}{\sqrt{N_\mu}}.\]
The total error on the oscillation probability has to account also for systematic
effects. The main systematic effect comes from the knowledge of the flux.
Without the near detector, this is given in first approximation by the 
statistical error on the number of electron neutrinos detected in the far
detector. Other sources
of systematics are corrections due to the knowledge of
the muon polarization, and from 
uncertainties on electron and muon identification efficiencies:
\[\frac{\Delta P}{P}=\frac{\Delta N_\mu}{N_\mu} \oplus \frac{\Delta N_e}{N_e} \oplus S\]
The contribution of the first two components of the total error, of pure
statistical nature, are plotted in Figure \ref{fig:errel} as a function of the
integrated flux. Even in the most favorable case, this error never gets below
2\%, and it is assumed that the other sources of systematics, indicated by S
in the above expression, are controlled to a better level, so they are 
neglected in the total error estimation.\par
Being of only statistical nature, the total error decreases as the square 
root of the total flux. If a near detector is built, the statistical error on
the number of $\nu_e$ detected becomes negligible, but then systematic effects
due to the extrapolation from the near to the far detector become 
relevant.\par
The error also shows a dependence on the inverse
of the square of the beam energy, because of the combined effect
of the increased neutrino cross section and of the larger number of 
neutrinos reaching the far detector as an effect of the Lorentz boost. 
Since, as seen before,
the oscillation probability decreases with the square of the beam energy,
at first approximation the sensitivity of the experiment is independent on the
energy.\par 
From the curves shown, it is possible to see that for the low energy beam
the oscillation probability is equal to the $1 \sigma$ error in the case of
$10^{22}$ muons for values of $\Delta m^2$ around $4\times 10^{-4} eV^2$.\par
\begin{figure}
  \begin{center}\mbox{
   \epsfig{file=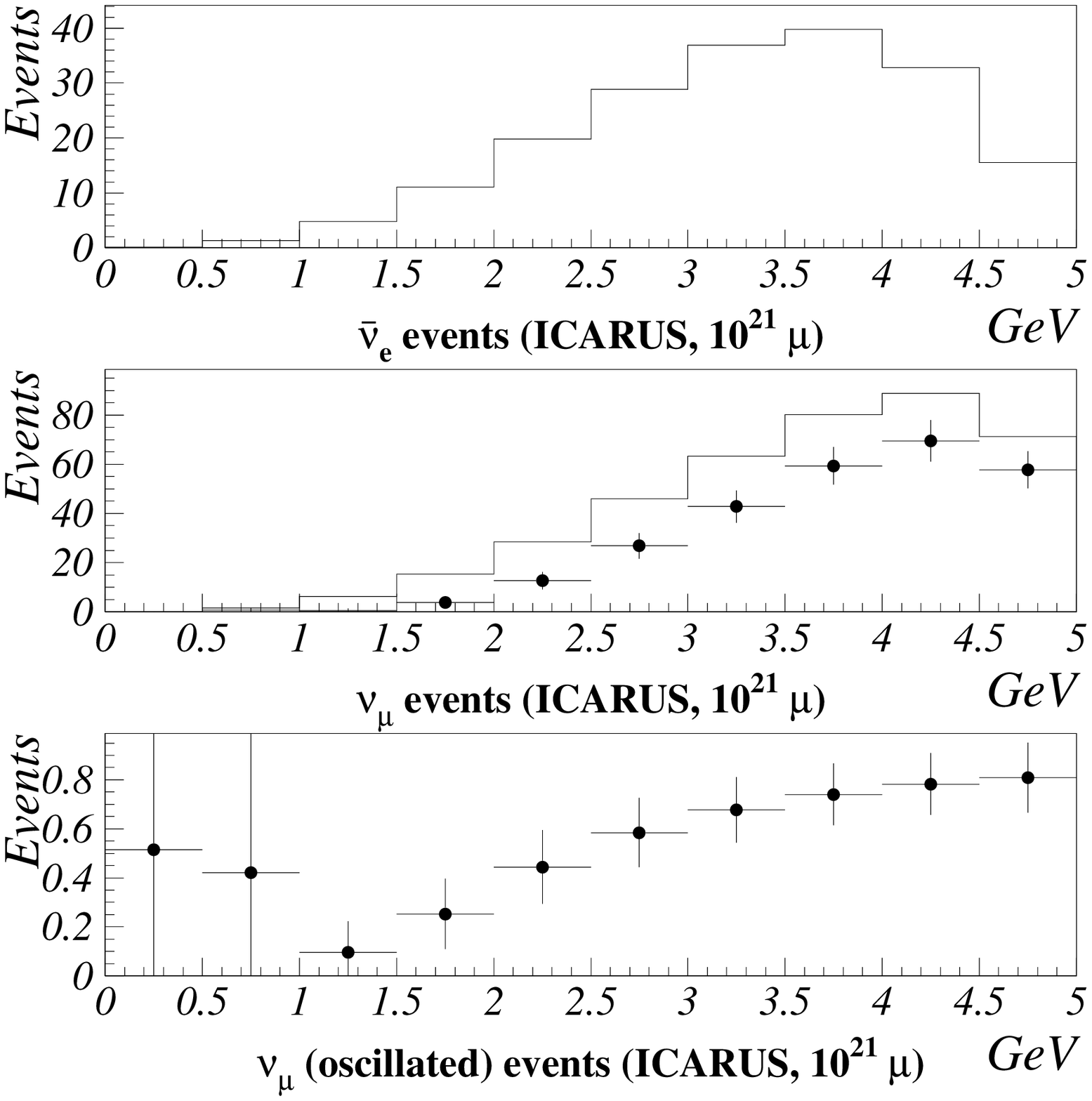,bb=0 120 567 687,width=.95\linewidth}
}
  \end{center}
  \caption{Energy spectra of detected $\bar{\nu_e}$ [upper plot], 
non-oscillated (solid line) and oscillated $\nu_\mu$ (dots)[middle plot] and the ratio of the two
[lower plot] for $10^{21}$ muons of 5.5 GeV and $\Delta m^2=2.2\times 10^{-3}$ eV$^2$}
  \label{fig:osce-3}
\end{figure}
\begin{figure}
  \begin{center}\mbox{
   \epsfig{file=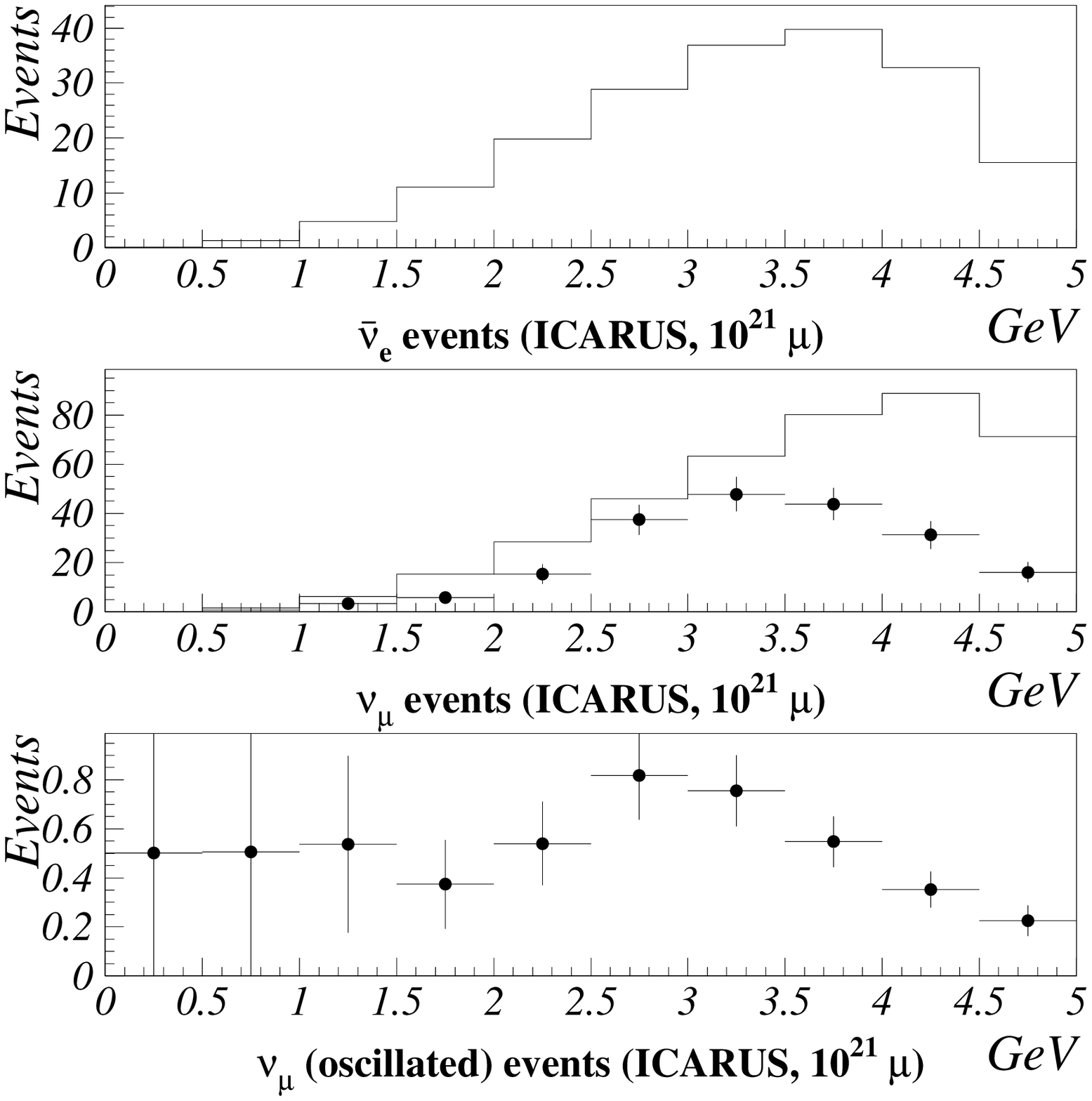,bb=0 120 567 687,width=.95\linewidth}
}
  \end{center}
  \caption{Energy spectra of detected $\bar{\nu_e}$ [upper plot], 
non-oscillated (solid line) and oscillated $\nu_\mu$ (dots) [middle plot] and the ratio of the two
[lower plot] for $10^{21}$ muons of 5.5 GeV and $\Delta m^2=1.0 \times 
10^{-2}$ eV$^2$}
  \label{fig:osce-2}
\end{figure}

The above discussion is based on the fact that at low oscillation probabilities
the oscillation occurs in the low energy part of the spectrum, where statistics
is limited.
At higher values of $\Delta m^2$, however, more information can be extracted 
from the energy distribution of the detected
neutrinos, in addition to the total flux, since in this case the characteristic
oscillation pattern can be seen. As an example, the spectra of $\bar{\nu_e}$,
$\nu_\mu$ without oscillations, $\nu_\mu$ with oscillations and their ratio 
are shown in Figures \ref{fig:osce-3} and \ref{fig:osce-2} assuming value of
$\Delta m^2$ of $2.2 \times 10^{-3} eV^2$
and $1.0 \times 10^{-2} eV^2$, and
considering the smearing due to the energy resolution of the ICARUS detector. 
The oscillation patterns are clearly visible.\par
\begin{figure}
  \begin{center}\mbox{
   \epsfig{file=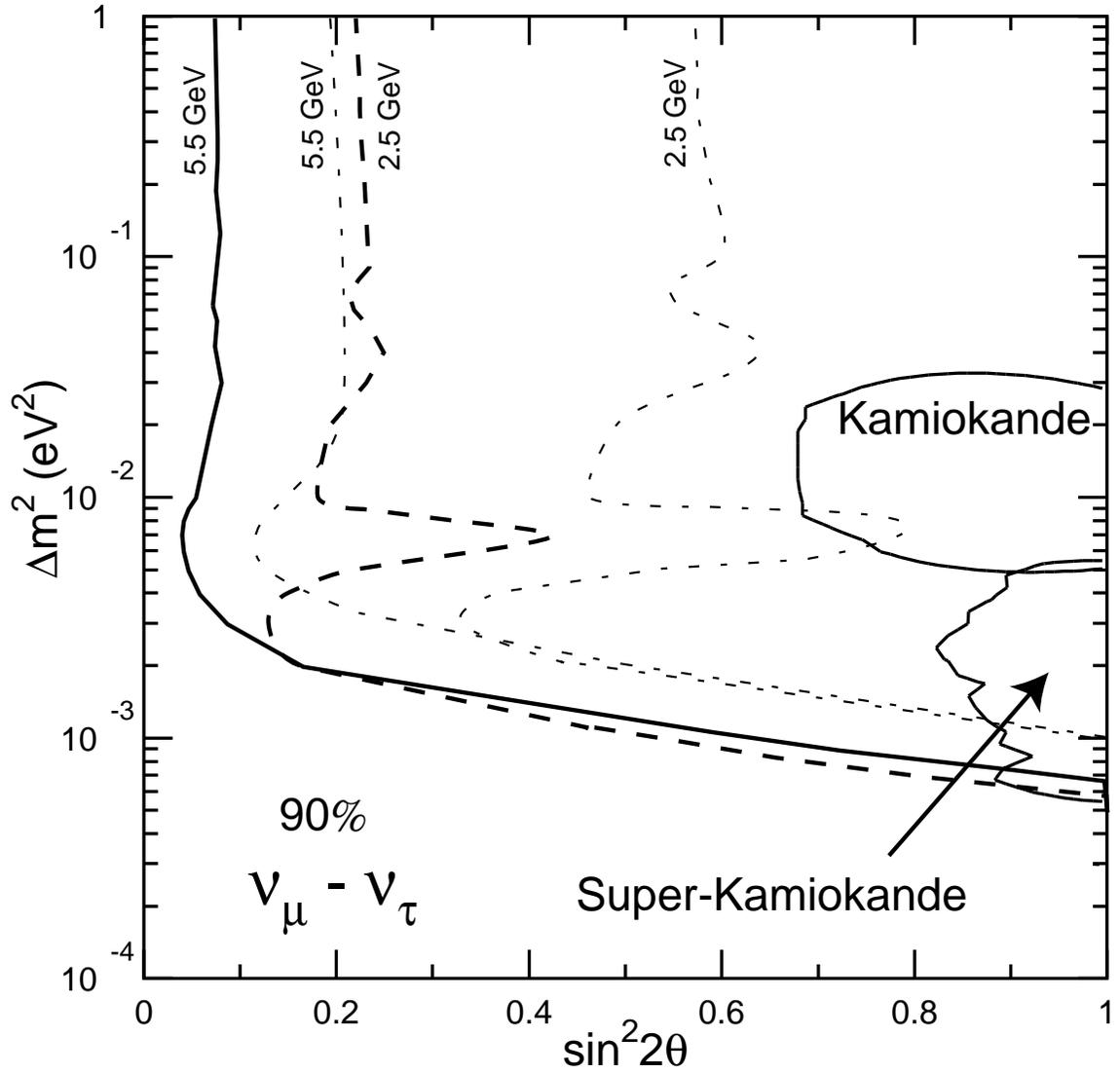,width=.95\linewidth}
}
  \end{center}
  \caption{90\% C.L contours for a disappearance experiment with a total flux of $10^{21}$ and $10^{22}$ muons, compared to the SuperKamiokande 90\% C.L.
contour. Full line is the limit for 5.5 GeV muon beam and $10^{22}$ muons,
dashed bold line is for 2.5 GeV beam and $10^{22}$ muons, dashed line is for
5.5 GeV beam and $10^{21}$ muons, dot-dashed line is for 2.5 GeV beam and
$10^{21}$ muons.}
  \label{fig:cont}
\end{figure}

All considerations done so far have been made in the assumption of maximal 
mixing.
When smaller mixing angles are also considered, two-dimensional oscillation
contours can be extracted. 
For the final result we consider the two cases of integrated fluxes:
$10^{21}$ and $10^{22}$ muons. The 90\% C.L. contours
are shown in Figure \ref{fig:cont} for a muon energy of 2.5 and 5.5 GeV. 
It can be seen that, while in the ``low'' intensity scenario
the covered area reaches values of $\Delta m^2$ of about $10^{-3} eV^2$, for 
the high intensity case the sensitivity goes to about $6\times10^{-4} eV^2$, 
thus covering at 90\% C.L. all the range indicated by the atmospheric neutrino 
results.
\par
\section{Conclusions}
The SuperKamiokande result, allowing the possibility of very small 
$\Delta m^2$ for  neutrino oscillations, pose strong challenges to the experiments
willing to explore this region of the parameter space. The method presented
here is based on disappearance of $\nu_\mu$ coming from a muon beam. 
Given the layout
of the accelerator, a near position could be placed on the surface,
opposite with respect to the far laboratory. Since a built-in control sample 
of $\bar{\nu_e}$ coming from the same muons can be exploited for normalization,
the experiment can however be performed without requiring a near detector, 
and in this configuration it has the 
potentiality of reaching in a clean way the low-$\Delta m^2$ region suggested
by the atmospheric neutrino results.



\end{document}